\documentclass[11pt]{article}
\usepackage{graphicx}
\usepackage{amssymb,amsmath}
\usepackage[sort&compress,numbers]{natbib}

\begin{document}

\title{
\textbf{Correlation functions, universal ratios and Goldstone mode singularities
 in $n$--vector models}
}

\author{J. Kaupu\v{z}s$^{1,2}$ 
\thanks{E--mail: \texttt{kaupuzs@latnet.lv}} \hspace{1ex}, 
R. V. N. Melnik$^3$, J. Rim\v{s}\=ans$^{1,2}$ \\
$^1$Institute of Mathematics and Computer Science, University of Latvia\\
29 Rai\c{n}a Boulevard, LV--1459 Riga, Latvia \\
$^2$ Institute of Mathematical Sciences and Information Technologies, \\
University of Liepaja, 14 Liela Street, Liepaja LV--3401, Latvia \\
$^3$Wilfrid Laurier University, Waterloo, Ontario, Canada, N2L 3C5}

\maketitle

\begin{abstract}
Correlation functions in the $O(n)$ models below the critical temperature 
are considered. Based on Monte Carlo (MC) data, 
we confirm the fact stated earlier by Engels and Vogt, that the transverse two--plane correlation function
of the $O(4)$ model for lattice sizes about $L=120$ and small 
external fields $h$ is very well described by a Gaussian approximation. However, we 
show that fits of not lower quality are provided by certain non--Gaussian
approximation. We have also tested larger lattice sizes, up to $L=512$. The Fourier--transformed
transverse and longitudinal two--point correlation functions
have Goldstone mode singularities in the thermodynamic limit at $k \to 0$ and $h=+0$, i.~e., 
$G_{\perp}({\bf k}) \simeq a k^{-\lambda_{\perp}}$ and 
$G_{\parallel}({\bf k}) \simeq b k^{-\lambda_{\parallel}}$, respectively.
Here $a$ and $b$ are the amplitudes, $k= \mid {\bf k} \mid$ is the magnitude of the wave vector ${\bf k}$. 
The exponents $\lambda_{\perp}$, $\lambda_{\parallel}$ and the 
ratio $b M^2/a^2$, where $M$ is the spontaneous
magnetization, are universal according to the GFD (grouping of Feynman diagrams)
approach. Here we find that the universality
follows also from the standard (Gaussian) theory, yielding
$b M^2/a^2 = (n-1)/16$. Our MC estimates of this ratio are $0.06 \pm 0.01$ for $n=2$, $0.17 \pm 0.01$ for $n=4$ 
and $0.498 \pm 0.010$ for $n=10$. According to these and our earlier MC results, the 
asymptotic behavior and Goldstone mode singularities are not exactly 
described by the standard theory. This is expected from the GFD theory. 
We have found appropriate analytic approximations for $G_{\perp}({\bf k})$ and $G_{\parallel}({\bf k})$, 
well fitting the simulation data for small $k$. We have used them to test the
Patashinski--Pokrovski relation and have found that it holds approximately.
\end{abstract}



\maketitle

\section{Introduction}
\label{intro}

The $n$--component vector--spin models (called also $n$--vector models or $O(n)$ models),  
have attracted significant interest in recent decades as the models, where the so--called Goldstone 
mode singularities are observed. The Hamiltonian of the $n$--vector model $\mathcal{H}$ is given by
\begin{equation}
\frac{\mathcal{H}}{T}=-\beta \left( \sum\limits_{\langle i j \rangle}
{\bf s}_i {\bf s}_j + \sum_i {\bf h s}_i \right) \;,
\end{equation} 
where $T$ is temperature, ${\bf s}_i \equiv {\bf s}({\bf x}_i)$ is the 
$n$--component vector of unit 
length, i.~e., the spin variable of the $i$--th lattice site with coordinate ${\bf x}_i$, 
$\beta$ is the coupling constant, and ${\bf h}$ is the external field. The summation takes place over
all nearest neighbors in the lattice. Periodic boundary conditions are considered here.

In the thermodynamic limit below the critical temperature (at $\beta>\beta_c$),
the magnetization $M(h)$ (where $h= \mid {\bf h} \mid$), the Fourier--transformed transverse ($G_{\perp}({\bf k})$)
and longitudinal ($G_{\parallel}({\bf k})$) two--point correlation functions 
exhibit Goldstone mode power--law singularities:
\begin{align}
&M(h) - M(+0) \propto h^{\rho} \quad \mbox{at} \quad h \to 0 \;, \label{M} \\ 
&G_{\perp}({\bf k}) = a \, k^{-\lambda_{\perp}} \;\; \mbox{at} \;\; h=+0 \;\; \mbox{and} \;\; k \to 0 \;, \label{a} \\
&G_{\parallel}({\bf k}) = b \, k^{-\lambda_{\parallel}} \quad \mbox{at} \;\; h=+0 \;\; \mbox{and} \;\; k \to 0 \label{b} 
\end{align}
with certain exponents $\rho$, $\lambda_{\perp}$, $\lambda_{\parallel}$ and
the amplitudes $a$, $b$ of the Fourier--transformed two--point correlation functions.

In a series of theoretical works (e.~g.,~\cite{Law1,Law2,HL,Tu,SH78,ABDS99,Dupuis}), it has been claimed
that the exponents in (\ref{M}) -- (\ref{b}) are exactly $\rho =1/2$ at $d=3$, $\lambda_{\perp}=2$ and $\lambda_{\parallel}=4-d$, 
where $d$ is the spatial dimensionality $2 < d < 4$. These theoretical approaches are further 
referred here as the standard theory.  
Several MC simulations have been performed earlier~\cite{DHNN,EM,EHMS,EV} to verify the compatibility of MC data
with standard--theoretical expressions, where the exponents are fixed.
In recent years, we have performed a series of accurate MC simulations~\cite{KMR08,KMR10,K2012,KMR12}
for remarkably larger lattices than previously were available, with an aim to evaluate the exponents in~(\ref{M}) -- (\ref{b}). 
Some deviations from the standard--theoretical values have been observed, in agreement with an
alternative theoretical approach, known as the GFD (grouping of Feynman diagrams) theory~\cite{K2010},
where the relations $d/2 < \lambda_{\perp} < 2$, $\lambda_{\parallel} = 2 \lambda_{\perp} - d$ and
$\rho = (d/\lambda_{\perp})-1$ have been found for $2<d<4$. 

Here we focus on the relations, which have not been tested in the previous MC studies (see Sec.~\ref{sec:corrf}).
In particular, the two--plane correlation function, studied in~\cite{EV}, is re-examined in Sec.~\ref{sec:fitstwo}.
Furthermore, we have also evaluated in Sec.~\ref{sec:ratios} the universal ratio $b M^2/a^2$ for $n=2,4,10$ and 
have compared the MC estimates with the values calculated here from the standard theory. Finally, in Sec.~\ref{sec:analap} 
we have proposed and tested certain analytical approximations for the two--point correlation functions, and 
in Sec.~\ref{sec:PP} have tested the Patashinski--Pokrovski relation (PP relation).

\section{Correlation functions}
\label{sec:corrf}

In presence of an external field ${\bf h}$, the longitudinal (parallel to  ${\bf h}$)
and the transverse (perpendicular to ${\bf h}$) spin components   
have to be distinguished. The Fourier--transformed longitudinal and transverse two--point
correlation functions are 
\begin{equation}
G_i({\bf k}) = \sum\limits_{\bf x} \grave{G}_i({\bf x}) \, e^{-i{\bf kx}} \;, 
\label{eq:corfur}
\end{equation}
where $i=1$ refers to the longitudinal component and $i=2, \ldots, n$ --- to the transverse ones.
Here
\begin{equation}
\grave{G}_i({\bf x}) = \langle s_i({\bf 0}) s_i({\bf x}) \rangle 
\label{eq:Gxdef}
\end{equation}
are the two--point correlation functions in the coordinate space.
(Note that the factor $N^{-1}$ in Eqs.~(1.2)--(1.3) of~\cite{KMR12} and~(28)--(29) of~\cite{K2012} 
is $N^{-1}=1$ according to the actual definitions~(\ref{eq:corfur})--(\ref{eq:Gxdef}).) 
  The inverse transform of~(\ref{eq:corfur}) is
\begin{equation}
\grave{G}_i({\bf x}) = L^{-3} \sum\limits_{\bf k} G_i({\bf k}) \, e^{i{\bf kx}} \;,
\label{eq:Gxtr}
\end{equation}
where $L$ is the linear lattice size.
In the following, the cumulant correlation function
 \begin{equation}
\tilde{G}_i({\bf x}) = \langle s_i({\bf 0}) s_i({\bf x}) \rangle 
- \langle  s_i({\bf 0}) \rangle \langle  s_i({\bf x}) \rangle 
\label{eq:Gxcum}
\end{equation}
will also be considered. It agrees with $\grave{G}_i({\bf x})$ for the transverse
components, whereas a nonzero constant contribution $\langle s_1 \rangle^2 =M^2$ is subtracted in the longitudinal case.

Following~\cite{EV}, the two--plane correlation function is defined as
\begin{equation}
 D_i(\tau) = L^2 \langle S_i(0) S_i(\tau) \rangle \;,
\end{equation}
where
\begin{equation}
 S_i(\tau) = L^{-2} \sum\limits_{x,y=0}^{L-1} s_i(x,y,\tau) 
\end{equation}
is the spin component $s_i$, which is averaged over the plane $z=\tau$, denoting
${\bf x} = (x,y,\tau)$.

Using the definition of $D_i(\tau)$, as well as the relations~(\ref{eq:Gxdef}) and~(\ref{eq:Gxtr}), 
we obtain
\begin{eqnarray}
 &&D_i(\tau) = \sum\limits_{x,y=0}^{L-1} \grave{G}_i(x,y,\tau) \nonumber \\ 
&&= L^{-3} \sum\limits_{m_1,m_2,m_3} \, \sum\limits_{x,y=0}^{L-1}
G_i \left(k_{m_1}, k_{m_2}, k_{m_3} \right) 
\exp \left[ \frac{2 \pi i}{L} (m_1 x + m_2 y + m_3 \tau) \right] \;,
\end{eqnarray}
where $\grave{G}_i(x,y,\tau) \equiv \grave{G}_i({\bf x})$ with ${\bf x} = (x,y,\tau)$
and $G_i \left(k_{m_1}, k_{m_2}, k_{m_3} \right) \equiv G_i({\bf k})$ with 
${\bf k} = (k_{m_1}, k_{m_2}, k_{m_3})$, $k_m=2 \pi m/L$.
The summation over indices $m_j$ goes from $1-L+[L/2]$ to $[L/2]$, where $[L/2]$ denotes the integer part of $L/2$.
According to the properties of the wave function $\exp \left[ \frac{2 \pi i}{L} (m_1 x + m_2 y + m_3 \tau) \right]$,
the summation over $x$ and $y$ gives vanishing result unless $m_1=m_2=0$. More precisely, it leads to the result
\begin{equation}
D_i(\tau) = \frac{1}{L} \sum\limits_{\ell}
G_i(k_{\ell}) \cos \left(k_{\ell} \tau  \right) \;,
\label{eq:D}
\end{equation}
where $G_i(k) \equiv G_i(0,0,k)$ is the Fourier--transformed two--point correlation function in the $\langle 100 \rangle$
crystallographic direction, and $k_{\ell} = 2 \pi \ell/L$ with $\ell \in [1-L+[L/2],[L/2]]$. 

The transverse two--plane correlation function in the Gaussian approximation
\begin{equation}
D_{\perp}^{\mathrm{Gauss}}(\tau) = \frac{\chi_{\perp}}{L} \sum\limits_{\ell}
\frac{m^2}{m^2+k_{\ell}^2} \cos \left(k_{\ell} \tau  \right)
\label{eq:DGauss}
\end{equation}
is obtained by setting the Gaussian two--point correlation function
\begin{equation}
 G_{\perp}^{\,\mathrm{Gauss}}(k) =  \frac{\chi_{\perp} \, m^2}{m^2+k^2}
\label{eq:GGauss}
\end{equation}
into~(\ref{eq:D})
instead of $G_i(k) = G_{\perp}(k)$ for $i \ge 2$. The parameter $m$ in~(\ref{eq:DGauss})--(\ref{eq:GGauss}) 
is interpreted as mass, and the known
relation $G_{\perp}(0) = \chi_{\perp}$ between the transverse correlation
function $G_{\perp}(0)$ and the transverse susceptibility $\chi_{\perp}$ is used here.

\begin{figure}
\begin{center}
\includegraphics[width=0.5\textwidth]{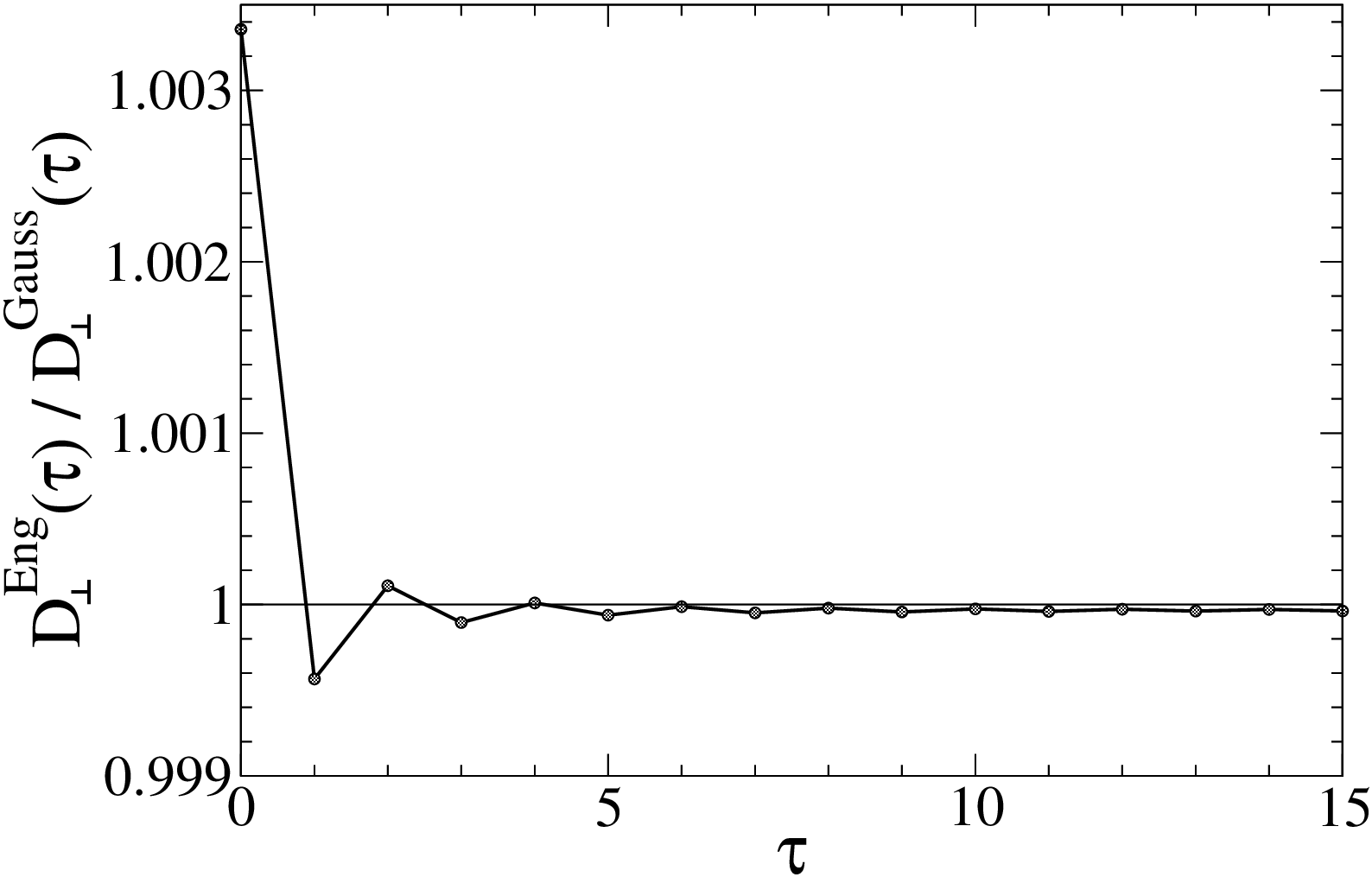}
\end{center}
\caption{The ratio 
$D_{\perp}^{\mathrm{Eng}}(\tau) / D_{\perp}^{\mathrm{Gauss}}(\tau)$
calculated from~(\ref{eq:DEng}) and~(\ref{eq:DGauss}) for $m=0.02$ and $L=120$
within $0 \le \tau \le 15$.}
\label{compare}
\end{figure}

A different formula has been proposed in~\cite{EV}, i.~e.,
\begin{equation}
D_{\perp}^{\mathrm{Eng}}(\tau) = \chi_{\perp} \tanh \left( \frac{m}{2} \right)
\frac{e^{-m \tau} + e^{-m (L-\tau)}}{1-e^{-m L}} \;.
\label{eq:DEng}
\end{equation}
Eq.~(\ref{eq:DEng}) is obtained assuming that $D_{\perp}(\tau)$
is proportional to $e^{-m \tau} + e^{-m (L-\tau)}$~\cite{EV}, as in the case of
the continuum limit, where the summation over wave vectors $2 \pi \ell/L$ runs
from $\ell =-\infty$ to $\ell = \infty$. Besides, the proportionality
coefficient is determined from the normalization condition
\begin{equation}
 \sum\limits_{\tau=0}^{L-1} D_{\perp}(\tau) = \chi_{\perp} \;.
\label{eq:normcond}
\end{equation}
Note that this condition is automatically satisfied in~(\ref{eq:D})
and~(\ref{eq:DGauss}) according to $G_{\perp}(0)=\chi_{\perp}$, 
since all terms cancel each other after the summation over $\tau$,
except only those with $k=0$.
It is clear that~(\ref{eq:DEng}) is not exactly consistent with~(\ref{eq:DGauss}),
as it can be easily checked by writing down all terms in~(\ref{eq:DGauss}), e.~g.,
at $L=2$ (where only terms with $\ell = 0$ and $\ell = 1$ appear).
However, the difference appears to be rather small for large $L$ and small $m$.
In Fig.~\ref{compare}, we have shown the ratio 
$D_{\perp}^{\mathrm{Eng}}(\tau) / D_{\perp}^{\mathrm{Gauss}}(\tau)$ for
the lattice size $L=120$ simulated in~\cite{EV} at a typical value of mass $m=0.02$ 
considered there. As we can see, the largest difference, about $0.3 \%$, appears at $\tau=0$.
It turns out that~(\ref{eq:DEng}) slightly better fits the simulation data around $\tau=0$,
although the quality of the overall fit is practically the same for~(\ref{eq:DGauss}) and~(\ref{eq:DEng}).
Since our aim is to test the consistency with the Gaussian spin--wave theory rather than to find
a very good fit formula for $D_{\perp}(\tau)$, we have used~(\ref{eq:DGauss}). 

It is interesting to mention that  $D_{\perp}^{\mathrm{Eng}}(\tau)$ corresponds to certain approximation
for $G_i(k)=G_{\perp}(k)$ in~(\ref{eq:D}), i.~e., to
\begin{equation}
 G^*_{\perp}(k) = \frac{2}{m} \tanh \left(\frac{m}{2} \right) 
\sum\limits_{j=-\infty}^{\infty} G_{\perp}^{\,\mathrm{Gauss}}(k+2 \pi j) \;.
\label{eq:D*}
\end{equation}
Indeed, an infinite sum 
over all integer values of $\ell$ is obtained when inserting the approximation~(\ref{eq:D*})
for $G_{\perp}(k)$ into~(\ref{eq:D}), yielding~(\ref{eq:DEng}) (see~\cite{EV} for
treatment of such sums). The correct normalization
is ensured here by the factor $\frac{2}{m} \tanh \left(\frac{m}{2} \right)$ in~(\ref{eq:D*}). 

The fact that~(\ref{eq:DEng}) better than~(\ref{eq:DGauss}) fits the simulation data for $D_{\perp}(\tau)$ near
$\tau =0$ can be understood based on~(\ref{eq:D*}). The data points for $G_{\perp}(k)$ lie on a smooth 
curve with a minimum at $k=\pi$ --- see Fig.~5 in~\cite{KMR10}. It is consistent with the fact that
$G_i(k) = G_i(2 \pi -k)$ holds and the data lie on a curve having no singularity at $k=\pi$.
The Gaussian approximation~(\ref{eq:GGauss}) does not have, but~(\ref{eq:D*}) does have this property.
Since it refers to large--$k$ behavior, the most significant difference between~(\ref{eq:DGauss}) and~(\ref{eq:DEng})
appears at small $\tau$ values.

\section{Fits of the two--plane correlation function}
\label{sec:fitstwo}

The two--point correlation functions $G_i(k)$ for $n=2,4,10$ have been extracted from MC
simulations by a modified Wolff cluster algorithm in our earlier works~\cite{KMR10,K2012,KMR12}. 
According to~(\ref{eq:D}), it allows us to evaluate also the two--plane correlation functions and compare the
results and conclusions with those of~\cite{EV}. In this case, it is meaningful to
determine the $G_{\perp}(0)$ value from the relation
\begin{equation}
G_{\perp}(0) = \chi_{\perp} = \frac{M(h)}{\beta h} \;,
\label{eq:Gchi}
\end{equation}  
which holds owing to the rotational symmetry of the model.
The statistical error for $\chi_{\perp}$, calculated as $M(h)/(\beta h)$, is much
smaller than that for $\chi_{\perp}^*=G_{\perp}(0)$, calculated
from the common formulas for $G_{\perp}(k)$ in~\cite{KMR10},
although the agreement within the error bars is expected
according to~(\ref{eq:Gchi}). 

We have calculated $D_{\perp}(\tau)$ (equal to $D_i(\tau)$ for $i \ge 2$) 
from~(\ref{eq:D}) and have fit the results to the Gaussian form~(\ref{eq:DGauss})
with $\chi_{\perp}$ being determined directly from simulations as $M(h)/(\beta h)$.
In this case, the only fit parameter is $m$. Our fit results for $m$, together
with the above discussed values of $\chi_{\perp}$ and $\chi_{\perp}^*$ for $O(n)$
models with $n=2,4,10$ are collected in Tabs.~\ref{fitGauss2} to~\ref{fitGauss10}.
The results for different lattice sizes $L$ at the smallest $h$ values in our simulations are 
shown here, providing also the $\chi^2/\mbox{d.o.f.}$ 
($\chi^2$ of the fit per degree of freedom) values, characterizing the quality of the fits.
A comparison between $\chi_{\perp}$ and $\chi_{\perp}^*$ for the $O(4)$ model
has been provided already in~\cite{KMR10}. In distinction from~\cite{KMR10},
here we do not use extra runs for  $\chi_{\perp}$, i.~e., both
quantities are extracted from the same simulation runs. 

We have found it convenient to split
any simulation run in $110$ bins, each including about $7.7 \times 10^{5}/L$ cluster
algorithm steps, discarding first $10$ bins for equilibration~\cite{KMR10}.
The statistical error of a quantity $X$ is evaluated by the jackknife 
method~\cite{MC} as $\sqrt{ \sum_i (X-X_i)^2}$, where $X_i$ is the $X$ value,
obtained by omitting the $i$-th bin. Here the bin-averages are considered
as statistically independent (or almost independent) quantities. It is well justified, since the
number of MC steps of one bin is much larger than that of the autocorrelation time. 
We have verified it by checking that the estimated statistical errors are 
practically the same when twice larger bins are used. The discarded $10$ bins
comprise a remarkable fraction of a simulation. It
ensures a very accurate equilibration. We have verified it by comparing the
estimates extracted from separate parts of a simulation.
The statistical errors in $G_{\perp}(k)$ at different $k$ values are correlated, since
$G_{\perp}(k)$ is measured simultaneously for all $k$. Hence, the 
statistical errors in $D_{\perp}(\tau)$ are correlated, as well.

\begin{table}
\caption{The estimates of transverse susceptibility  
$\chi_{\perp}$ and $\chi_{\perp}^*$
(see text) and the fit parameter $m$ (mass) in~(\ref{eq:DGauss}) for the $O(2)$ model at $\beta=0.55$ and $h=0.00021875$
depending on the lattice size $L$.
The values of $\chi^2/\mbox{d.o.f.}$ of the fit are given in the last column.}
\label{fitGauss2}
\begin{center}
\begin{tabular}{|c|c|c|c|c|}
\hline
\rule[-2.5mm]{0mm}{7mm}
$L$ & $m$            & $\chi_{\perp}$ & $\chi_{\perp}^*$ & $\chi^2/\mbox{d.o.f.}$  \\ \hline
512 & 0.01714(41)    & 5254.762(75)   & 4645(387)        &   1.24              \\
384 & 0.01681(44)    & 5254.765(79)   & 5236(368)	 &   0.67              \\ 
256 & 0.01690(25)    & 5254.22(19)    & 5846(445)        &   1.27              \\
128 & 0.01717(16)    & 5245.68(67)    & 5321(212)        &   0.90              \\
64  & 0.016738(76)   & 5142.7(1.5)    & 5230(60)         &   0.53              \\ \hline
\end{tabular}
\end{center}
\end{table}

\begin{table}
\caption{The susceptibility estimates 
$\chi_{\perp}$ and $\chi_{\perp}^*$,
the fit parameter $m$ in~(\ref{eq:DGauss}), and the $\chi^2/\mbox{d.o.f.}$ values of the fit for the $O(4)$ 
model at $\beta=1.1$ and $h=0.0003125$ vs size $L$.}
\label{fitGauss4}
\begin{center}
\begin{tabular}{|c|c|c|c|c|}
\hline
\rule[-2.5mm]{0mm}{7mm}
$L$ & $m$            & $\chi_{\perp}$ & $\chi_{\perp}^*$ & $\chi^2/\mbox{d.o.f.}$  \\ \hline
350 & 0.02381(41)    & 1422.831(40)   & 1449(75)	 &   1.01              \\ 
256 & 0.02423(34)    & 1422.775(60)   & 1435(64)         &   0.28              \\
128 & 0.02398(16)    & 1420.98(18)    & 1404(42)         &   0.40              \\
64  & 0.024041(92)   & 1389.21(57)    & 1386(16)         &   0.51              \\ \hline
\end{tabular}
\end{center}
\end{table}

\begin{table}
\caption{The susceptibility estimates 
$\chi_{\perp}$ and $\chi_{\perp}^*$,
the fit parameter $m$ in~(\ref{eq:DGauss}), and the $\chi^2/\mbox{d.o.f.}$ values of the fit
 for the $O(10)$ model at $\beta=3$ and $h=0.00021875$ vs size $L$.}
\label{fitGauss10}
\begin{center}
\begin{tabular}{|c|c|c|c|c|}
\hline
\rule[-2.5mm]{0mm}{7mm}
$L$ & $m$            & $\chi_{\perp}$ & $\chi_{\perp}^*$ & $\chi^2/\mbox{d.o.f.}$  \\ \hline
350 & 0.02042(28)     & 719.464(24)    & 732(28)          &   4.63              \\
256 & 0.02147(27)    & 719.468(36)    & 665(25)	         &   1.28              \\ 
192 & 0.02086(19)    & 719.176(58)    & 744(24)          &   0.29              \\
128 & 0.02103(12)    & 717.47(14)     & 715(19)          &   0.22              \\
64  & 0.021530(71)   & 690.07(36)     & 694.8(6.2)       &   0.38              \\ \hline
\end{tabular}
\end{center}
\end{table}

\begin{figure}
\begin{center}
\includegraphics[width=0.6\textwidth]{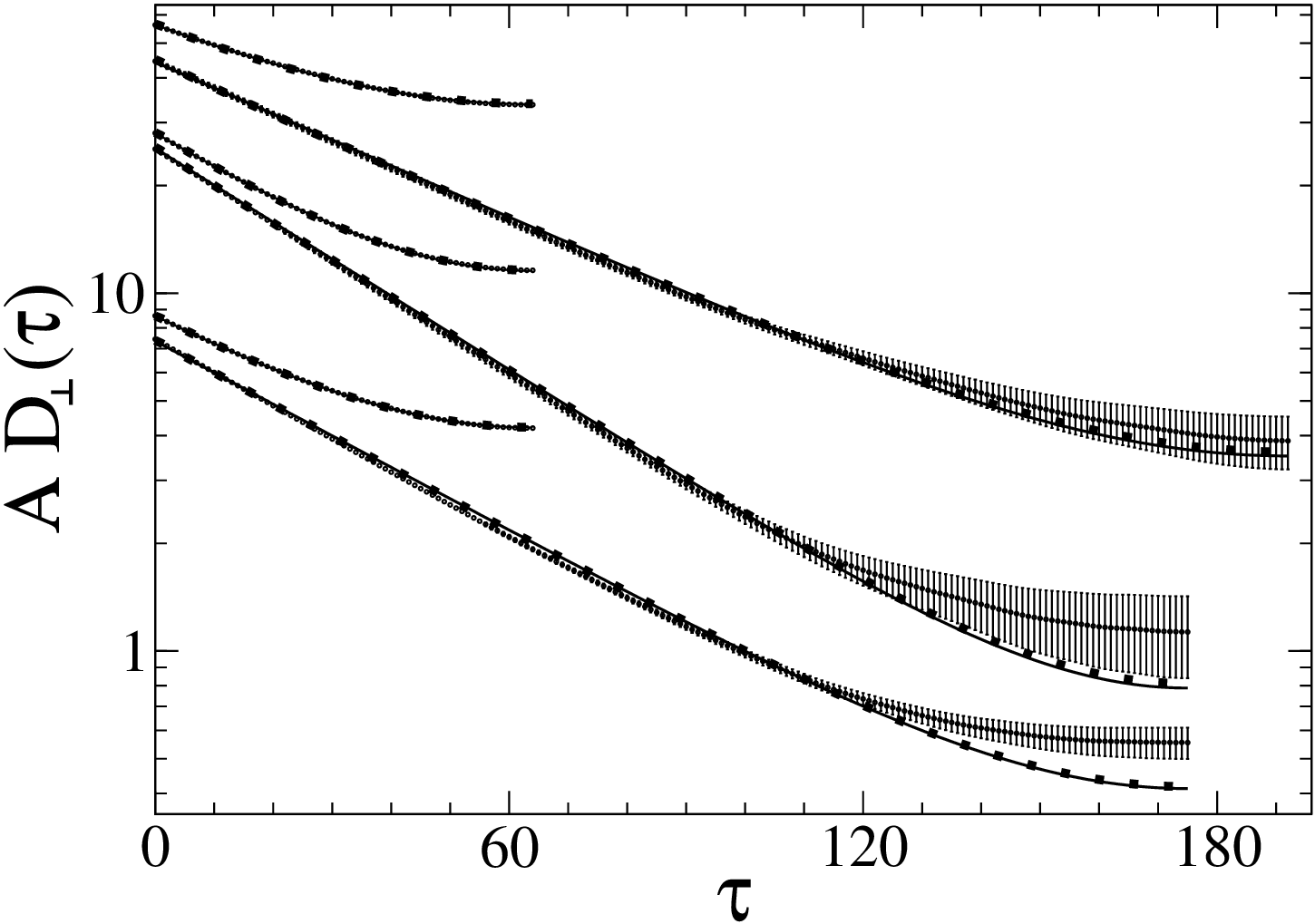}
\end{center}
\caption{
$A D_{\perp}(\tau)$ vs $\tau$ plots of the transverse two--plane correlation 
function $D_{\perp}(\tau)$ in the $O(n)$ model. The results are shown for $n=2$ 
(upper plots, $A=1$, $\beta=0.55$, $h=0.00021875$, $L=128$ and $384$), 
$n=4$ (middle plots, $A=1.5$, $\beta=1.1$, $h=0.0003125$, $L=128$ and $350$), 
and $n=10$ (lower plots, $A=1$, $\beta=3$, $h=0.00021875$, $L=128$ and $350$). 
The curves extend to $\tau \le L/2$. 
The solid lines are fits to the Gaussian form~(\ref{eq:DGauss}), the dotted lines ---
fits to~(\ref{eq:D}) with the approximation~(\ref{eq:apper0}) for $G_{\perp}(k)$,
where $\lambda_{\perp} =1.929, 1.955, 1.972$ for $n=2,4,10$, respectively.}
\label{planecorr}
\end{figure}

The fit curves for $L=128$ and for a larger size, $L=350$ or $L=384$, are plotted
by solid lines in Fig.~\ref{planecorr}. The fits look perfect for $L=128$ (short curves).
In such a way, we confirm the results of~\cite{EV}, where perfect fits
for a similar size $L=120$ have been obtained in the case of $n=4$.
However, our fits are less perfect for larger sizes (longer curves). In the cases of $n=2$ and $n=4$,
the discrepancies about one standard error can be explained
by correlated statistical errors in the $D_{\perp}(\tau)$ data. 
However, the deviations of the data points from the fit curve are remarkably larger 
for $n=10$ and $L=350$, as it can be seen from
the lower plots in Fig.~\ref{planecorr}, as well as from the relatively large $\chi^2/\mbox{d.o.f.}$ value
$4.63$ in this case -- see Tab.~\ref{fitGauss10}.

The authors of~\cite{EV} tend to interpret the very good fits of $D_{\perp}(\tau)$ to
the ansatz~(\ref{eq:DEng}) for the $O(4)$ model at $L=120$ 
as an evidence that the model is essentially Gaussian, implying that the exponent in~(\ref{a})
is $\lambda_{\perp}=2$. Recall that~(\ref{eq:DEng}) is not exactly the same as~(\ref{eq:DGauss}), but 
the difference is insignificant, as discussed in Sec.~\ref{sec:corrf}.  

A serious reason why, in our opinion, the argument of~\cite{EV} cannot be regarded as a real proof
or evidence that $\lambda_{\perp}=2$ really lies with the fact that practically the same or even better fit
is provided by a non-Gaussian approximation of the form 
\begin{equation}
 G_{\perp}(k) \approx \chi_{\perp} \left( \frac{\tilde{a}}{\tilde{a}+k^2} \right)^{\lambda_{\perp}/2}
\label{eq:apper0}
\end{equation}
for the transverse two--point correlation function in~(\ref{eq:D}) with certain values of $\lambda_{\perp}<2$.
This approximation will be discussed in detail in Sec~\ref{sec:analap}. Here we only note that it is consistent
with the Gaussian form at $\tilde{a}=m^2$ and $\lambda_{\perp}=2$, as well as 
with the general power--law asymptotic $a k^{-\lambda_{\perp}}$ at $h \to 0$ under an appropriate choice of 
$\tilde{a} = \tilde{a}(h)$. We have considered $\tilde{a}$ as the only fit parameter at a fixed exponent
$\lambda_{\perp}=1.955$, consistent with the estimation for $n=4$ in~\cite{KMR10}. The $\chi^2/\mbox{d.o.f.}$
value of the resulting $D_{\perp}(\tau)$ fit for the $O(4)$ model at $L=128$ is $0.23$. It is smaller
than the value $0.4$ of the Gaussian fit (see Tab.~\ref{fitGauss4}) and even smaller than the value
$0.36$ of the fit to~(\ref{eq:DEng}). We have considered also the non-Gaussian fits with 
$\lambda_{\perp}=1.929$ for $n=2$ and $\lambda_{\perp}=1.972$ for $n=10$, as consistent with our
estimation of the exponents in~\cite{KMR08,KMR12}. The non-Gaussian fits are shown by dotted lines
in Fig.~\ref{planecorr}. As we can see, the Gaussian and non-Gaussian fit curves lie almost on top
of each other. It means the analysis of the two--plane correlation functions hardly can 
give any serious evidence about the exponent $\lambda_{\perp}$. 

It refers also to the spectral analysis of~\cite{EV}, where the transverse spectral 
function $\bar{A}(\omega)$ is defined as the solution of the integral equation
\begin{equation}
 D_{\perp}(\tau) = \int\limits_0^{\infty} \bar{A}(\omega) \bar{K}(\omega,\tau) d \omega \;,
\end{equation}
with the kernel
\begin{equation}
 \bar{K}(\omega,\tau) = \tanh \left(\frac{\omega}{2} \right) \, 
\frac{e^{-\omega \tau} + e^{-\omega (L-\tau)}}{1 - e^{-\omega L}} \;.
\end{equation}
According to~\cite{EV}, the solution is $\bar{A}(\omega) \approx \chi_{\perp} \delta(\omega-m)$.
Numerically we never get the delta function, so that practically the spectrum consists of a sharp peak at $\omega = m$.
In fact, $\bar{A}(\omega) \approx \chi_{\perp} \delta(\omega-m)$ means only that 
$D_{\perp}(\tau) \approx \chi_{\perp} \bar{K}(m,\tau)$ holds as a good approximation.
According to the discussed here consistency of different fits, the latter is possible
if the small--$k$ asymptotic of $G_{\perp}(k)$ is given either by~(\ref{eq:GGauss}), or by~(\ref{eq:D*}), or 
by~(\ref{eq:apper0}) with appropriate value of $\lambda_{\perp}<2$. Thus, no clear conclusion 
concerning $\lambda_{\perp}$ can be drawn here.

In fact, we need a direct estimation of the exponents, as in our papers~\cite{KMR08,KMR10,K2012,KMR12},
to judge seriously whether or not the asymptotic behavior of correlation
functions and related quantities are Gaussian. Our estimation
suggests that these are non-Gaussian.

Deviations of the simulated data points from the lower fit curves in Fig.~\ref{planecorr}
are practically the same for $\lambda_{\perp}=2$ and $\lambda_{\perp}=1.972$ in~(\ref{eq:apper0})
(solid and dotted lines). Hence, if these deviations are not caused mainly by correlated and larger than usually 
statistical fluctuations, then one has to conclude that corrections to the 
form~(\ref{eq:apper0}) are relevant in this case.

\section{Universal ratios}
\label{sec:ratios}

The ratio $b M^2/a^2$, composed of the amplitudes $a$, $b$ and magnetization $M=M(+0)$ in~(\ref{M}) -- (\ref{b}),
is universal according to~\cite{K2010}. The ratio $B M^2/A^2$, where $A$ and $B$ are the corresponding amplitudes
of the real--space correlation functions, can be easily related to $b M^2/a^2$. 
In the thermodynamic limit $L \to \infty$  for large $x = \mid {\bf x} \mid$ we have
\begin{equation}
 \tilde{G}_i({\bf x}) = \tilde{G}_i(x) = \frac{1}{x} \, \frac{1}{2 \pi^2} \int\limits_0^{\infty}
 f(k) \,  k \, G_i(k) \sin(kx) dk 
\label{eq:Gx}
\end{equation}
in three dimensions, where $f(k)$ is the cut--off function, which we choose as
\begin{equation}
f(k) = \frac{1}{1 + (k/\Lambda)^4} \;,
\label{eq:cutoff}
\end{equation}
where $\Lambda$ is a constant.
This result is obtained by subtracting the constant contribution from~(\ref{eq:Gxtr}), provided by ${\bf k =0}$,
and replacing the remaining sum over ${\bf k}$ by the corresponding integral, taking into account that
the correlation functions are asymptotically (at $x \to \infty$ or $k \to 0$) isotropic in the thermodynamic
limit. Here we use a smooth cut--off in the ${\bf k}$--space, which can be chosen quite arbitrary
(however, ensuring the convergence of the integral), since only 
the small--$k$ contribution is relevant for the large--$x$ behavior. Hence, (\ref{eq:Gx}) is valid 
with $G_i(k) = G_i(0,0,k)$.  

As everywhere in this paper, $i=1$ can be replaced with ``$\parallel$'' and $i \ge 2$ --- with ``$\perp$''.
The asymptotic of $\tilde{G}_{\perp}(x) = A x^{\lambda_{\perp} -3}$ at $x \to \infty$, corresponding to
 $G_{\perp}(k) = a k^{-\lambda_{\perp}}$ at $k \to 0$,
as well as $\tilde{G}_{\parallel}(x) = B x^{\lambda_{\parallel} -3}$ at $x \to \infty$, corresponding to
 $G_{\parallel}(k) = b k^{-\lambda_{\parallel}}$ at $k \to 0$,
 can be easily calculated from~(\ref{eq:Gx}), using the known relation~\cite{Erdel,Fedoruk}
\begin{equation}
 \int\limits_0^{\infty} k^{\alpha -1} f(k) \, \sin(kx) \, dk  \sim
x^{-\alpha} \, f(0) \, \Gamma(\alpha) \, \sin \left(\frac{\pi \alpha}{2} \right)
\label{eq:erdei}
\end{equation}
for $\alpha>0$ (the Erd\'elyi Lemma~\cite{Erdel} applied to our particular case).
It yields  
\begin{equation}
 \frac{b M^2}{a^2} = \frac{BM^2}{A^2} \, \frac{1}{2\pi^2} \, 
\frac{\Gamma^2(\eta^*) \sin^2 \left(\frac{\pi}{2} \eta^* \right)}{\Gamma(1+2 \eta^*) 
\sin \left(\frac{\pi}{2}(1+2\eta^*) \right)}
\label{eq:ratios}
\end{equation}
for $\eta^* = 2-\lambda_{\perp}>0$ and $\lambda_{\parallel} = 2 \lambda_{\perp} -3$,
corresponding to the relations of the GFD theory at $d=3$~\cite{K2010}. 
The ratio $bM^2/a^2$ and, consequently, also $BM^2/A^2$ are universal in this theory.
The standard--theoretical case $\eta^*=0$ is recovered at $\eta^* \to 0$
in~(\ref{eq:ratios}), as it can be checked by direct calculations.
In this case, the usage of~(\ref{eq:erdei}) at $\alpha=0$ is avoided,
applying the known relation between the $1/(k^2+m^2)$ asymptotic in ${\bf k}$--space
and the $e^{-mx}/(4 \pi x)$ asymptotic in ${\bf x}$--space
and taking the limit $m \to 0$. Thus, we obtain
\begin{equation}
 \left( \frac{b M^2}{a^2} \right)_{\mathrm{st}} = \frac{1}{8} \, \left( \frac{BM^2}{A^2} \right)_{\mathrm{st}} \;, 
\label{eq:ratiosst}
\end{equation}
where the subscript ``st'' indicates that the quantity is calculated within the standard theory.

One of the cornerstones of the standard theory is the Patashinski--Pokrovski (PP) relation
(see, e.~g.,~\cite{EV} and references therein) 
\begin{equation}
\tilde{G}_{\parallel}({\bf x}) = \frac{n-1}{2M^2} \, \tilde{G}_{\perp}^2({\bf x}) \;.
\label{eq:PP} 
\end{equation}
It is supposed that~(\ref{eq:PP}) holds in the ordered phase in the thermodynamic limit for large distances,
i.~e., ${\bf x}$ can be replaced by $x = \mid {\bf x} \mid$ here. 
According to~(\ref{eq:PP}) and~(\ref{eq:ratiosst}), we have
\begin{eqnarray}
\left( \frac{BM^2}{A^2} \right)_{\mathrm{st}} &=& \frac{n-1}{2} \;, \label{eq:st1} \\
\left( \frac{b M^2}{a^2} \right)_{\mathrm{st}} &=& \frac{n-1}{16} \label{eq:st2} \;.
\end{eqnarray}
 It turns out that these amplitude ratios can be precisely calculated in the
standard theory, and they appear to be universal, as predicted by the GFD theory.
The accuracy of the standard theory can be checked by comparing~(\ref{eq:st1})--(\ref{eq:st2})
with Monte Carlo estimates.

According to the relation $\lambda_{\parallel} = 2 \lambda_{\perp} -d$, which holds in the GFD 
theory~\cite{K2010} and also in the standard theory (where $\lambda_{\perp}=2$ and 
$\lambda_{\parallel}=4-d$), in 3D case we have
\begin{equation}
 \frac{b M^2}{a^2} = \lim\limits_{k \to 0} R(k) \;,
\end{equation}
where the quantity
\begin{equation}
 R(k) = \frac{M^2 G_{\parallel}(k)}{k^3 G_{\perp}^2(k)}
\label{eq:R}
\end{equation}
is calculated in the thermodynamic limit at $h=+0$.
In order to estimate $R(0)$ in this limit, we consider appropriate
range of $k$ values, i.~e. $k > k^*$, for small fields $h$ and large system sizes $L$,
where the finite--size effects are very small or practically negligible and the finite--$h$ effects are also 
small. Then, we extrapolate the $R(k)$ plots to $k=0$ at several $h$ values to find  
the required asymptotic value of $R(0)=b M^2/a^2$. 
Such analysis has been already performed in~\cite{KMR12} for the $O(4)$ model at $\beta=1.1$ and $\beta=1.2$, with
an aim to test the universality of $b M^2/a^2$ predicted in~\cite{K2010}. It has been confirmed, providing an estimate
$b M^2/a^2 = 0.17 \pm 0.01$ valid for both values of $\beta$. Now we can see from~(\ref{eq:st2}) 
that this estimate is slightly smaller than the standard--theoretical value $3/16 = 0.1875$. 

\begin{figure}
\begin{center}
\includegraphics[width=0.6\textwidth]{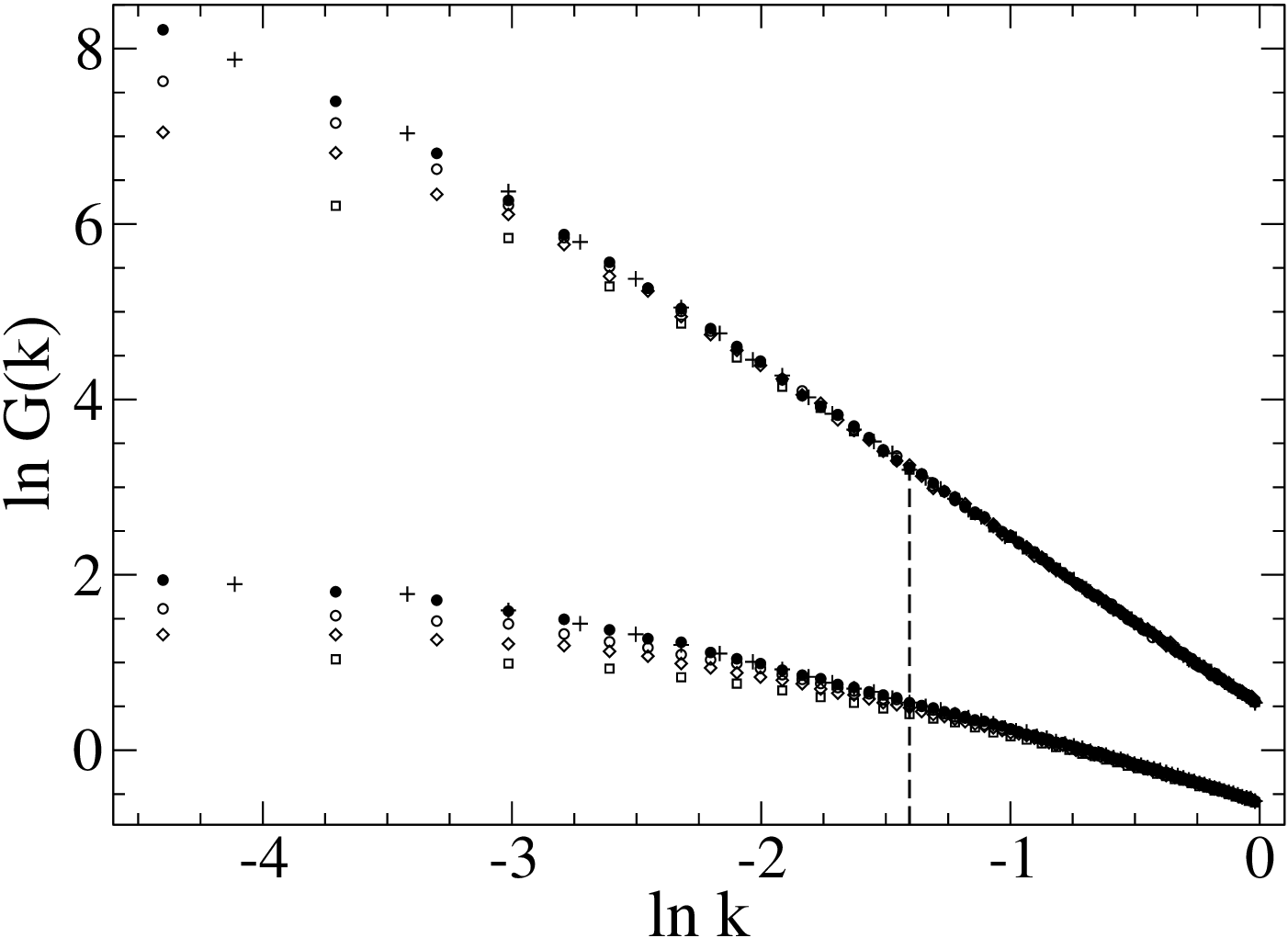}
\end{center}
\caption{Log--log plots of the transverse (top) and the longitudinal (bottom) Fourier--transformed two--point
correlation functions in the $O(2)$ model (shown for $k <1$ at $\beta=0.55$) at 
$h=h_{\mathrm{min}}=0.00021875$ and $L=512$ (solid circles), 
$h=h_{\mathrm{min}}$ and $L=384$ (pluses), $h=2 h_{\mathrm{min}}$ and $L=512$ (empty circles), 
$h=4 h_{\mathrm{min}}$ and $L=512$ (diamonds), $h=8 h_{\mathrm{min}}$ and $L=256$ (squares).
The vertical dashed line indicates the lower border of the $k$ interval  
used in the analysis of the amplitude ratio $b M^2 /a^2$.}
\label{corr}
\end{figure}

Here we consider the cases $n=2$ and $n=10$. The choice of the $k$--interval for the $O(2)$ model is illustrated
in Fig.~\ref{corr}, where we can see that the finite--size and finite--$h$ effects are
small for $k \ge k_{20}$ with  $k_{\ell}=2 \pi \ell/512$, as indicated by a dashed line.
Similarly, we have found that the region $k \ge k_{15}$ with $k_{\ell}=2 \pi \ell/350$ is appropriate
for our analysis at $n=10$. The plots similar to those in Fig.~\ref{corr} for $n=10$ are given 
in~\cite{KMR12} (see Figs.~1 and 2 there). 
We will start our estimation just with $n=10$, since the results are more precise and convincing in this case. 
\begin{figure}
\begin{center}
\includegraphics[width=0.6\textwidth]{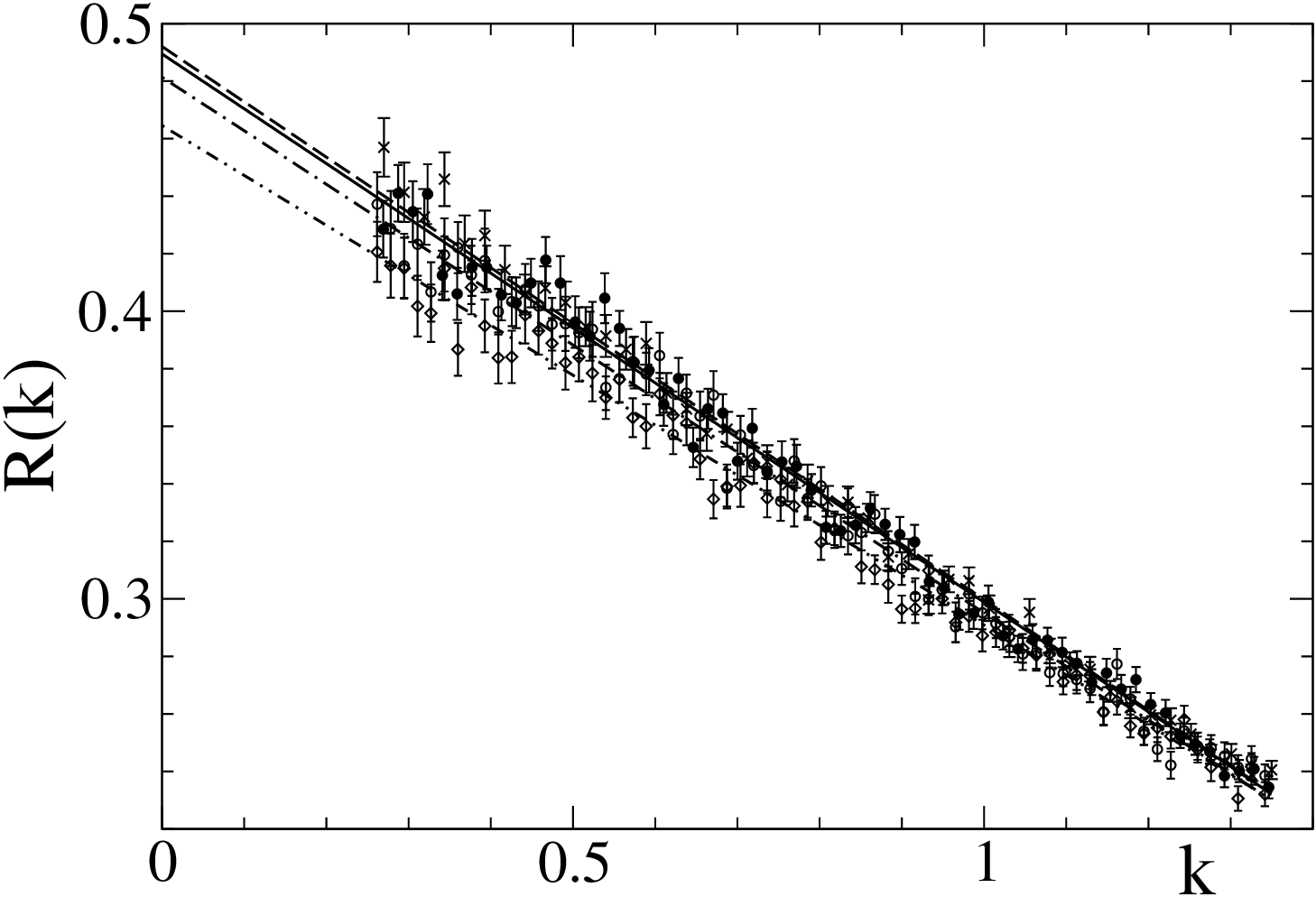}
\end{center}
\caption{The ratio $R(k)$~(\ref{eq:R}) in the $O(10)$ model at $\beta=3$. The results for 
$h=h_{\mathrm{min}}=0.00021875$ and $L=350$
(solid circles), $h=h_{\mathrm{min}}$ and $L=256$ (exes), $h=2 h_{\mathrm{min}}$ and $L=384$ (empty circles), 
as well as for $h=4 h_{\mathrm{min}}$ and $L=384$ (diamonds) are presented. The respective linear fits are shown by solid,
dashed, dot-dashed and dot-dot-dashed lines. The fit interval is $k_{15} \le k \le k_{75}$
with $k_{\ell}=2 \pi \ell/350$ for $L=350$ and similar in other cases.}
\label{ratio_10}
\end{figure}
According to the corrections to scaling of the standard theory, the correlation functions are expanded
in powers of $k^{4-d}$ and $k^{d-2}$~\cite{Law1,SH78}, i.~e., in powers of $k$ at small wave vectors in
three dimensions. It means that the ratio $R(k)$ is expected to be linear function of $k$ at $k \to 0$.
We indeed observe a very good linearity for the $O(10)$ model within $k_{15} \le k \le k_{75}$ 
($k_{\ell}= 2 \pi \ell /350$), as it can be seen from Fig.~\ref{ratio_10},
where the fit results for this or very similar intervals are shown at different fields $h$ and lattice sizes $L$. 
For the smallest $h$ value $h=h_{\mathrm{min}} = 0.00021875$, the linear fits give
$R(0) = 0.4895(27), 0.4920(27), 0.4936(26)$ and $0.5021(28)$ at $L=350, 256, 192$ and $128$, respectively. Hence we can judge
that the finite--size effects are practically negligible at $L=350$. The results for $h=2 h_{\mathrm{min}}$
and  $h=4 h_{\mathrm{min}}$ at $L=384$ are $R(0)=0.4814(27)$ and $R(0)=0.4646(28)$, respectively.
\begin{figure}
\begin{center}
\includegraphics[width=0.5\textwidth]{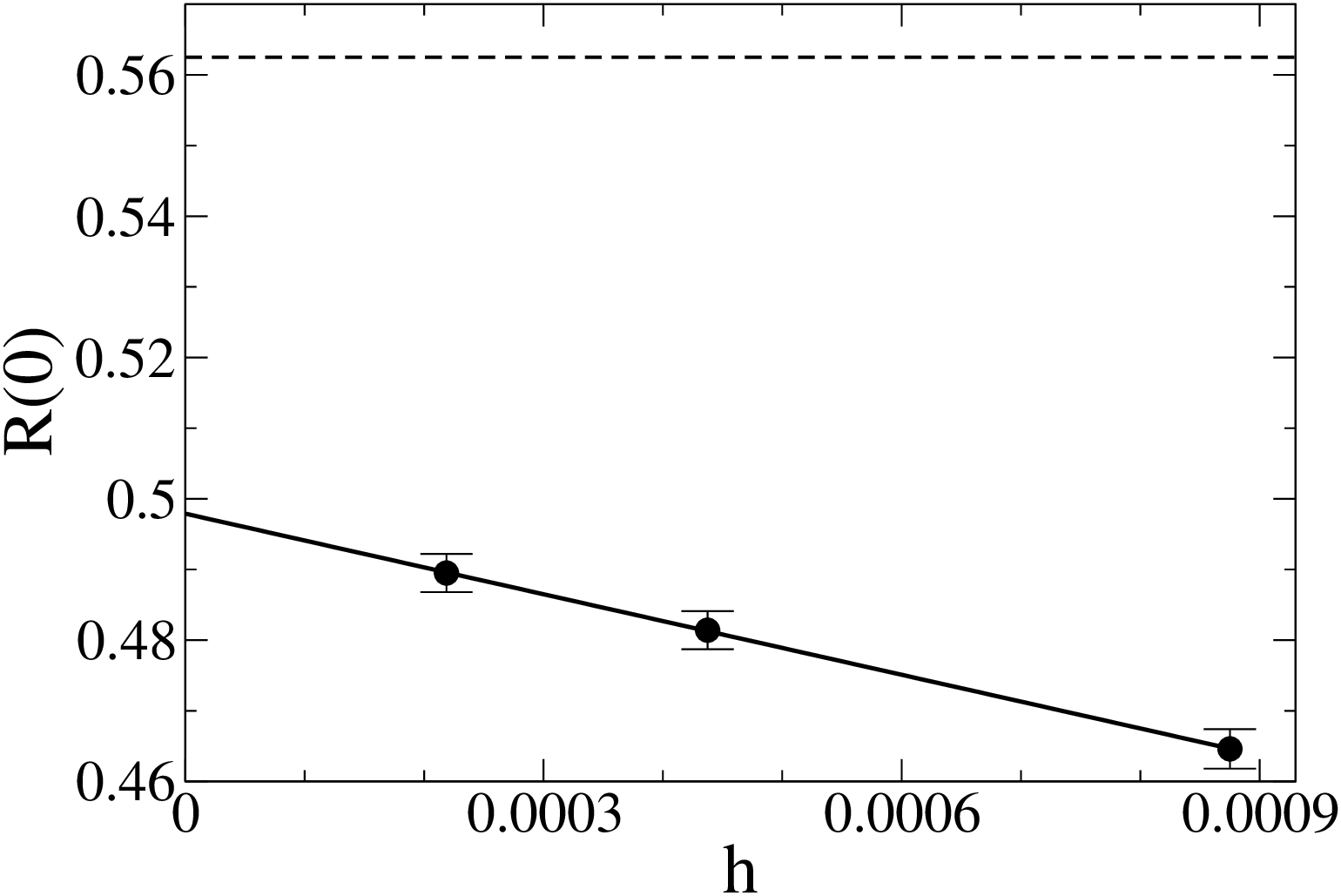}
\end{center}
\caption{The values of $R(0)$ (solid circles), evaluated from the fits in Fig.~\ref{ratio_10} 
depending on the external field $h$, taking the largest size $L$ for each $h$. The linear fit
gives an estimate $R(0) = 0.4979(33)$ for $h=+0$. The
standard--theoretical value $9/16 = 0.5625$ is indicated by dashed line.}
\label{R0}
\end{figure}
In Fig.~\ref{R0}, the $R(0)$ estimates for the largest sizes are shown depending on $h$.
The three data points almost precisely fit on a straight line, which gives the asymptotic
estimate $R(0)=0.4979(33)$ for $h=+0$. This fit is plausible from the point of view that
the $h$--dependence is, indeed, expected to be smooth (analytic) for a fixed interval 
of nonzero $k$ values, where $R(k)$ has been calculated. Note, however, that the indicated
here error bars $ \pm 0.0033$ include only the statistical error. A systematic error can arise from a
weak nonlinearity of the plot and also from finite--size effects, which seem to be
smaller than the statistical error bars in this case. Since the possible non-linearity
is not well controlled having only three data points, we have set remarkably larger
error bars $\pm 0.01$ for our final estimate $R(0) = b M^2/a^2= 0.498 \pm 0.010$.
This estimate shows a small, but very remarkable as compared to the error bars,
deviation from the standard--theoretical value~(\ref{eq:st2}) $bM^2/a^2= 9/16=0.5625$,
indicated in Fig.~\ref{R0} by a dashed line.

A similar estimation is performed here for the $O(2)$ model, with an essential difference
that the $R(k)$ plots appear to be rather non-linear, well fit to a parabola instead of a straight line.
\begin{figure}
\begin{center}
\includegraphics[width=0.6\textwidth]{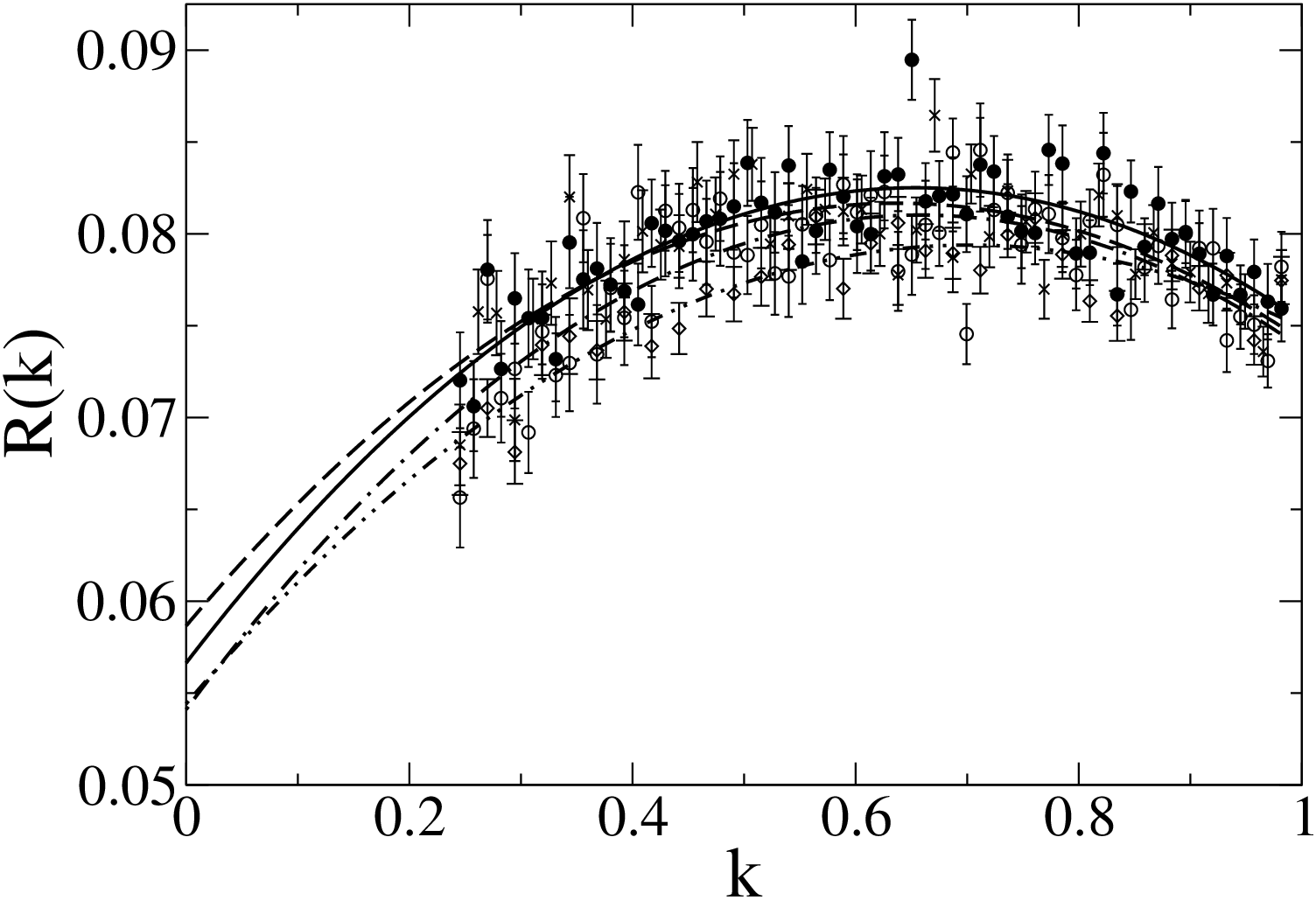}
\end{center}
\caption{The ratio $R(k)$~(\ref{eq:R}) in the $O(2)$ model at $\beta=0.55$. The results for 
$h= 2 h_{\mathrm{min}}=0.0004375$ and $L=512$
(solid circles), $h= 2 h_{\mathrm{min}}$ and $L=384$ (exes), $h=4 h_{\mathrm{min}}$ and $L=512$ (empty circles), 
and for $h=8 h_{\mathrm{min}}$ and $L=256$ (diamonds) are presented. The respective quadratic fits are shown by solid,
dashed, dot-dashed and dot-dot-dashed lines and yield
$R(0)=0.0566(22), 0.0587(26), 0.0541(25)$ and $0.0544(23)$.
The fit interval is $k_{20} \le k \le k_{80}$ with $k_{\ell}=2 \pi \ell/512$ for $L=512$ and similar in other cases.}
\label{ratio_2}
\end{figure}
Besides, in this case we have used the data for larger fields $h= 2 h_{\mathrm{min}}$,
$h= 4 h_{\mathrm{min}}$ and $h= 8 h_{\mathrm{min}}$, since the agreement between the results for
different lattice sizes at $h=  h_{\mathrm{min}}$ was not as good (although the estimate $R(0)=0.0590(24)$
at the largest size $L=512$ and $h= h_{\mathrm{min}}$, probably, is good).
In such a way, based on the fits shown in Fig.~\ref{ratio_2}, we have made a rough estimation 
$R(0) = b M^2/a^2 = 0.06 \pm 0.01$ for the $O(2)$ model. It agrees within the error bars with the standard--theoretical
value $1/16 = 0.0625$.

\section{Analytic approximations for 
$G_{\perp}(k)$ and $G_{\parallel}(k)$}
\label{sec:analap}

Let us now consider  the approximation~(\ref{eq:apper0}) in more detail. This approximation
does not uniquely follow from the theory in~\cite{K2010}, since the letter refers mainly to the case $h=+0$ in the 
thermodynamic limit. However, this approximation for non-zero $h$ together with an analogous one
for the longitudinal correlation function, i.~e.,   
\begin{eqnarray}
 G_{\perp}(k) &\approx& \chi_{\perp} \left( \frac{\tilde{a}(h)}{\tilde{a}(h)+k^2} \right)^{\lambda_{\perp}/2} \;,
\label{eq:apper} \\
 G_{\parallel}(k) &\approx& \chi_{\parallel} \left( \frac{\tilde{b}(h)}{\tilde{b}(h)+k^2} \right)^{\lambda_{\parallel}/2}
\label{eq:appar}
\end{eqnarray}
have the expected properties under appropriate choice of parameters $\tilde{a} = \tilde{a}(h)$ and $\tilde{b} = \tilde{b}(h)$.
The formulas~(\ref{eq:apper}) and~(\ref{eq:appar}) ensure that the correlation functions
can be expanded in powers of $k^2$ in vicinity of $k=0$ for any nonzero $h$. At the same time
they ensure the power--law asymptotic $G_{\perp}(k) = a \, k^{-\lambda_{\perp}}$
and $G_{\parallel}(k) = b \, k^{-\lambda_{\parallel}}$ at $h \to 0$ provided that 
$\tilde{a}(h) \sim \tilde{b}(h) \sim h^{2/\lambda_{\perp}}$ 
holds at $h \to 0$, taking into account the relations $\chi_{\perp} = M(h)/(\beta h)$ and 
$\chi_{\parallel} \sim h^{-\lambda_{\parallel}/\lambda_{\perp}}$. The latter one is true at $h \to 0$
according to Eq.~(9.25) in~\cite{K2010}. This behavior of $\tilde{a}(h)$ and $\tilde{b}(h)$
implies that $\xi_{\perp}(h) \sim \xi_{\parallel}(h) \sim h^{-1/\lambda_{\perp}}$ holds
for small $h$, where $\xi_{\perp}$ and $\xi_{\parallel}$ are the transverse and
the longitudinal correlation lengths.
Similar conclusion follows from the PP relation~(\ref{eq:PP}), i.~e.,
$\xi_{\perp}/\xi_{\parallel}=2$. However, according to~(\ref{eq:apper})--(\ref{eq:appar}), 
the ratio $\xi_{\perp}/\xi_{\parallel}$ is expected to be a constant at $h \to 0$, but not necessarily two.

Apparently, Eqs.~(\ref{eq:apper})--(\ref{eq:appar}) represent the simplest possible form
having the above discussed properties. Therefore, this form might be a very reasonable first 
approximation. Recall that the simulated quantities $G_{\perp}(k)$ and $G_{\parallel}(k)$ are the correlation
functions in the $\langle 100 \rangle$ crystallographic direction. However, since the two--point correlation
functions are isotropic at small $k$, the expressions in the right hand side of~(\ref{eq:apper}) and (\ref{eq:appar})
are generally meaningful approximations for $G_{\perp}({\bf k})$ and $G_{\parallel}({\bf k})$ with $k = \mid {\bf k} \mid$. 

In the following, we have considered $\tilde{a}(h)$ and $\tilde{b}(h)$, as well as the exponents
$\lambda_{\perp}$ and $\lambda_{\parallel}$ as fit parameters in~(\ref{eq:apper}) and~(\ref{eq:appar}).
In such a way, (\ref{eq:apper}) is consistent also with the standard theory if $\lambda_{\perp}=2$ holds
within the error bars. We have found that~(\ref{eq:apper}) fairly well fits our data for $O(n)$ models at various parameters 
within $k < 0.55$. The fit results are collected in Tab.~\ref{fitcortran}.  
\begin{table}
\caption{Parameters used in~(\ref{eq:apper}), $\tilde{a}(h)$
and $\lambda_{\perp}$ being evaluated from fits within $k<0.55$.}
\label{fitcortran}
\begin{center}
\begin{tabular}{|c|c|c|c|c|c|c|c|}
\hline
\rule[-2.5mm]{0mm}{7mm}
$n$ & $\beta$ & $10^4 h$ & $L$ & $\chi_{\perp}$ & $10^4 \tilde{a}(h)$ & $\lambda_{\perp}$ & $\chi^2/\mbox{d.o.f.}$  \\ \hline
2   & 0.55     & 2.1875  & 512 & 5254.762(75)   & 2.727(46)           & 1.9763(56)        & 0.72                    \\
2   & 0.55     & 4.375   & 512 & 2630.392(25)   & 5.713(83)           & 1.9892(53)        & 0.59                    \\ 
2   & 0.55     & 8.75    & 512 & 1317.3967(86)  & 11.30(19)           & 1.9835(71)        & 1.41                    \\
4   & 1.1      & 3.125   & 350 & 1422.831(40)   & 5.353(72)           & 1.9710(49)        & 1.60                    \\
4   & 1.1      & 4.375   & 350 & 1018.173(24)   & 7.70(11)            & 1.9810(58)        & 1.19                    \\  
4   & 1.2      & 4.375   & 350 & 1075.028(19)   & 6.854(82)           & 1.9863(46)        & 0.99                    \\ 
10  & 3        & 2.1875  & 350 & 719.464(24)    & 4.225(46)           & 1.9785(39)        & 1.46                    \\
10  & 3        & 4.375   & 384 & 361.3551(75)   & 8.556(96)           & 1.9824(45)        & 0.65                    \\  
10  & 3        & 8.75    & 384 & 181.8192(29)   & 16.98(20)           & 1.9797(53)        & 1.37                    \\  \hline    
\end{tabular}
\end{center}
\end{table}
The results of fits to~(\ref{eq:appar}) for the longitudinal two--point correlation function are presented
in Tab.~\ref{fitcorlong}. In this case, it is not always possible to fit well the data within $k<0.55$,
but the fits improve significantly (on average) for a narrower interval $k<0.28$.
\begin{table}
\caption{Parameters used in~(\ref{eq:appar}), 
$\tilde{b}(h)$
and $\lambda_{\parallel}$ being evaluated from fits within $k<k_{\mathrm{max}}$.}
\label{fitcorlong}
\begin{center}
\begin{tabular}{|c|c|c|c|c|c|c|c|c|}
\hline
\rule[-2.5mm]{0mm}{7mm}
$n$ & $\beta$ & $10^4 h$ & $L$ & $k_{\mathrm{max}}$ & $\chi_{\parallel}$ & $10^2 \tilde{b}(h)$ & $\lambda_{\parallel}$ & $\chi^2/\mbox{d.o.f.}$  \\ \hline
2   & 0.55     & 2.1875  & 512 & 0.28               & 7.62(25)           & 0.110(16)           & 0.736(12)        & 1.82     \\
2   & 0.55     & 2.1875  & 512 & 0.55               & 7.62(25)           & 0.123(14)           & 0.7599(46)       & 1.65     \\
2   & 0.55     & 4.375   & 512 & 0.28               & 5.29(15)           & 0.229(38)           & 0.688(19)        & 1.39     \\
2   & 0.55     & 4.375   & 512 & 0.55               & 5.29(15)           & 0.290(34)           & 0.7476(79)       & 2.27     \\
2   & 0.55     & 8.75    & 512 & 0.28               & 3.764(95)          & 0.70(13)            & 0.753(36)        & 0.88     \\
2   & 0.55     & 8.75    & 512 & 0.55               & 3.764(95)          & 0.701(83)           & 0.749(11)        & 0.78     \\
4   & 1.1      & 3.125   & 350 & 0.28               & 7.41(20)           & 0.276(35)           & 0.869(22)        & 1.49     \\
4   & 1.1      & 3.125   & 350 & 0.55               & 7.41(20)           & 0.370(32)           & 0.9707(92)       & 3.10     \\
4   & 1.1      & 4.375   & 350 & 0.28               & 6.36(17)           & 0.401(59)           & 0.888(29)        & 1.17     \\
4   & 1.1      & 4.375   & 350 & 0.55               & 6.36(17)           & 0.501(47)           & 0.973(12)        & 1.70     \\
4   & 1.2      & 4.375   & 350 & 0.28               & 4.27(14)           & 0.354(57)           & 0.875(27)        & 0.82     \\
4   & 1.2      & 4.375   & 350 & 0.55               & 4.27(14)           & 0.422(47)           & 0.936(12)        & 1.42     \\
10  & 3        & 2.1875  & 350 & 0.28               & 4.18(20)           & 0.210(44)           & 0.944(34)        & 1.60     \\
10  & 3        & 2.1875  & 350 & 0.55               & 4.18(20)           & 0.283(39)           & 1.052(14)        & 2.15     \\
10  & 3        & 4.375   & 384 & 0.28               & 2.624(98)          & 0.67(14)            & 1.038(61)        & 0.89     \\
10  & 3        & 4.375   & 384 & 0.55               & 2.624(98)          & 0.78(10)            & 1.103(21)        & 0.81     \\
10  & 3        & 8.75    & 384 & 0.28               & 1.920(72)          & 1.15(32)            & 1.020(95)        & 0.71     \\
10  & 3        & 8.75    & 384 & 0.55               & 1.920(72)          & 1.40(21)            & 1.129(33)        & 0.70     \\  \hline    
\end{tabular}
\end{center}
\end{table}

If we consider such fits as a method of estimation of the exponents $\lambda_{\perp}$ and $\lambda_{\parallel}$,
then it has certain advantage as compared to the estimations in~\cite{K2012,KMR12}, i.~e.,
 it is not necessary to discard the smallest $k$ values in order to ensure the smallness of the finite--$h$ effects.
However, a disadvantage is that no corrections to scaling are included in~(\ref{eq:apper})--(\ref{eq:appar}).
Therefore, the values reported in~\cite{KMR08,KMR10,K2012,KMR12} are preferable as asymptotic estimates.
The exponents in Tabs.~\ref{fitcortran} and~\ref{fitcorlong} slightly depend on the fit range, 
as well as on the field $h$. One can expect that they converge to the values of~\cite{KMR08,KMR10,K2012,KMR12}.
Indeed, at the smallest fields, the estimates of $\lambda_{\perp}$ for $O(4)$ and $O(10)$ models
in Tab.~\ref{fitcortran} are consistent with the corresponding values $1.955 \pm 0.020$ and $1.960(10)$ for $n=4$ 
and $1.9723(90)$ for $n=10$ reported in~\cite{KMR10,KMR12}. Moreover, in this case the longitudinal exponent 
$\lambda_{\parallel} = 2 \lambda_{\perp} - d$, calculated from these asymptotic estimates, is consistent
with $\lambda_{\parallel}$ for relatively small $k$ values ($k_{\mathrm{max}}=0.28$) at the smallest fields $h$ in 
Tab.~\ref{fitcorlong}. For the $O(2)$ model, the agreement between 
$\lambda_{\perp}=1.929(21)$, obtained in~\cite{KMR08} from~(\ref{M}) via scaling relation 
$\rho = (d/\lambda_{\perp})-1$~\cite{K2010}, 
and the smallest--$h$ estimate $\lambda_{\perp}=1.9763(56)$ in Tab.~\ref{fitcortran} is worse.
The exponent $\lambda_{\parallel}=0.736(12)$ in Tab.~\ref{fitcorlong} (at minimal $h$ and $k_{\mathrm{max}}=0.28$)
is somewhat smaller than the value $0.858(42)$, calculated from the scaling relation 
$\lambda_{\parallel} = 2 \lambda_{\perp} - d$~\cite{K2010} with $\lambda_{\perp}=1.929(21)$, 
although it agrees well with the direct estimation $\lambda_{\parallel}=0.69 \pm 0.10$ in~\cite{K2012}. 
The discrepancies indicate that corrections to scaling, including non-trivial ones of the GFD theory 
(discussed in~\cite{K2012,KMR12}), which have not been taken into account in the fitting procedures, 
are larger for the $O(2)$ model as compared to the $O(4)$ and $O(10)$ models. 

\begin{figure}
\begin{center}
\includegraphics[width=0.5\textwidth]{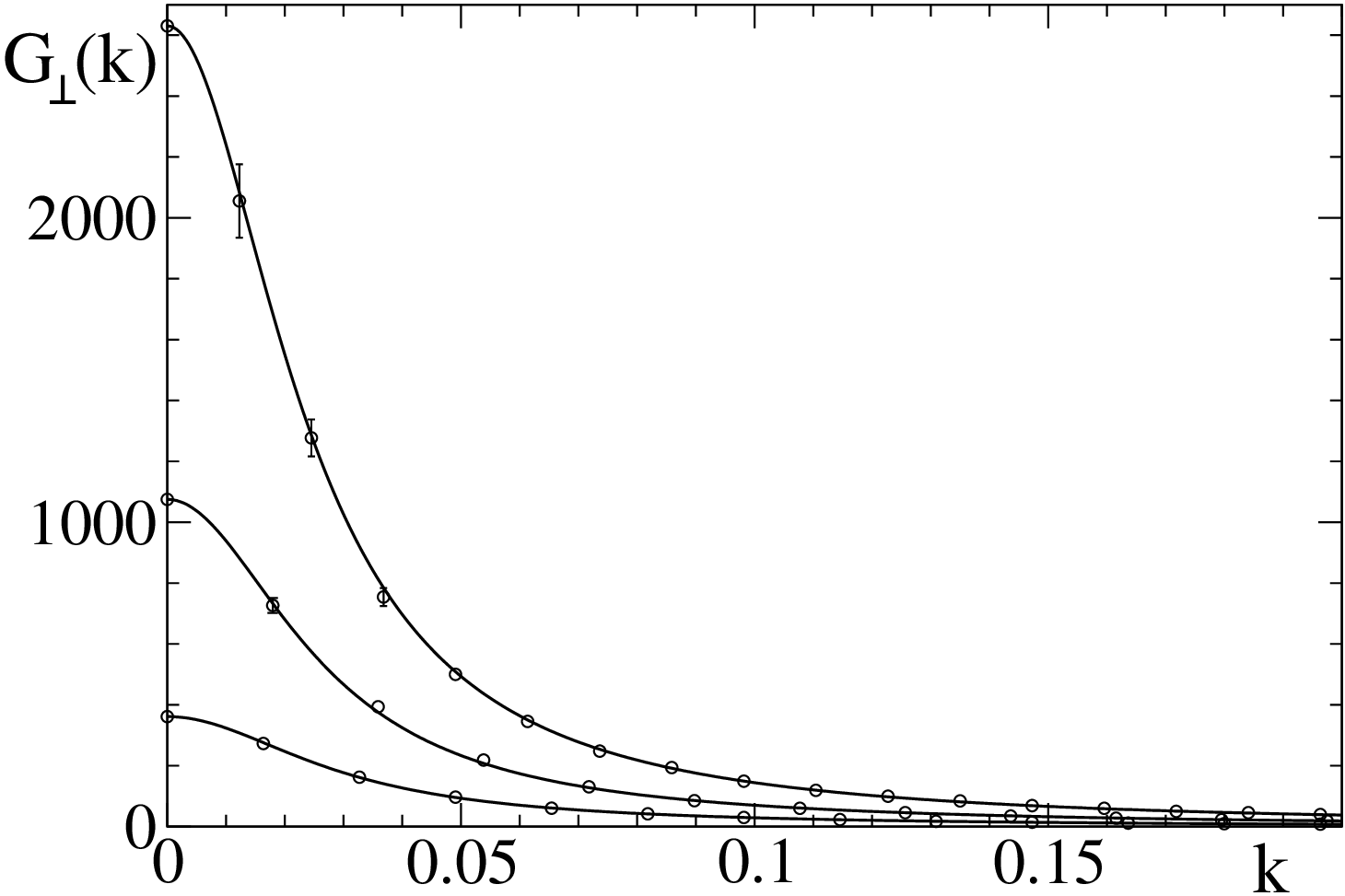}
\end{center}
\caption{The transverse two--point correlation function 
$G_{\perp}(k)$ at the value of external field $h=0.0004375$.
The results for the $O(2)$ model at $\beta=0.55$ and $L=512$ (top), $O(4)$ model at $\beta=1.2$ and $L=350$ (middle), 
and $O(10)$ model at $\beta=3$ and $L=384$ (bottom) are presented.
The simulated data points are shown by circles. The error bars are indicated,
where these are larger than the symbol size. Curves represent~(\ref{eq:apper}) with parameters
$\tilde{a}(h)$ and $\lambda_{\perp}$ evaluated from the fits within $k < 0.55$.}
\label{corrtra_fit}
\end{figure}

\begin{figure}
\begin{center}
\includegraphics[width=0.5\textwidth]{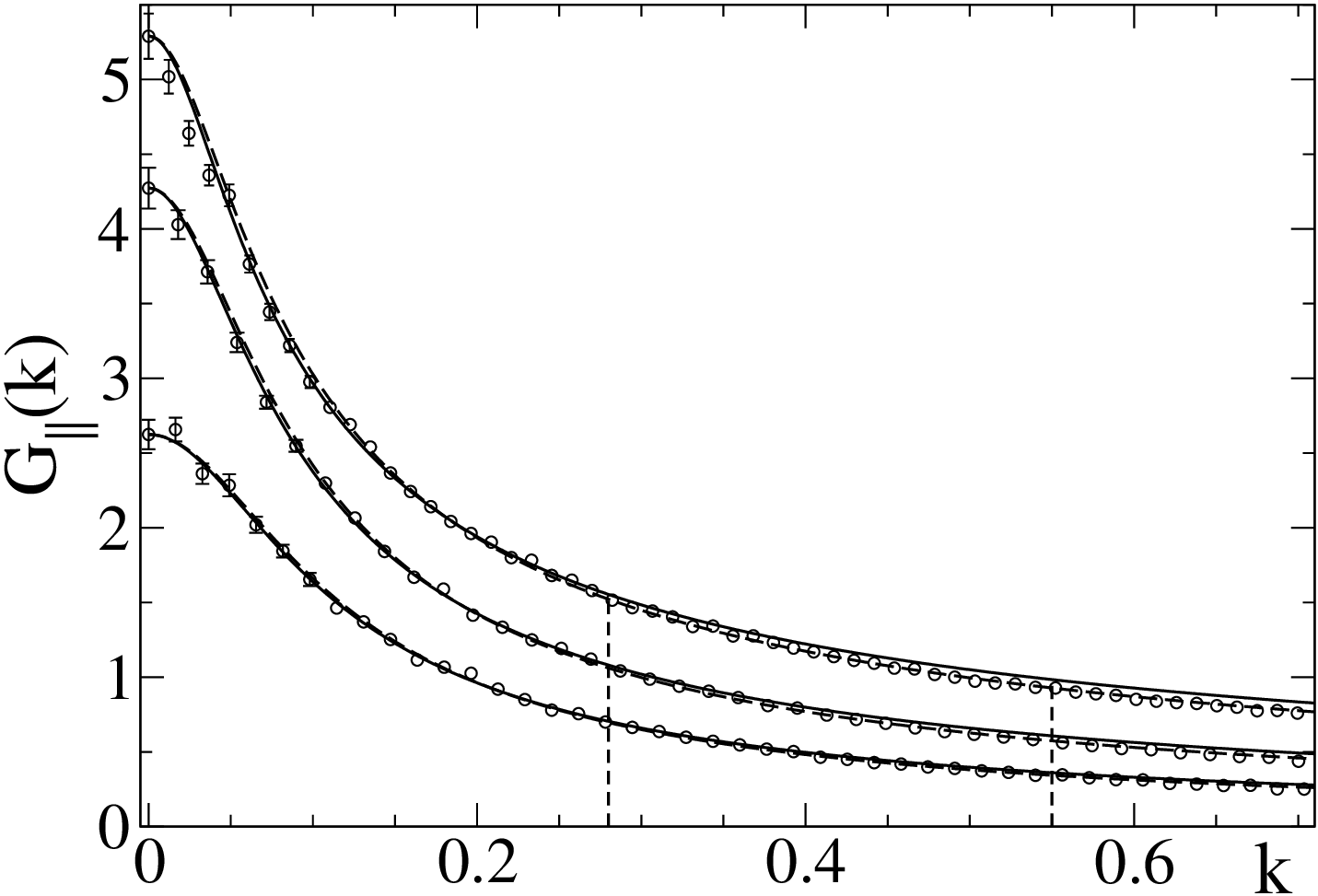}
\end{center}
\caption{The longitudinal two--point correlation function 
$G_{\parallel}(k)$
for the $O(2)$ model (top), $O(4)$ model (middle), 
and $O(10)$ model (bottom) at the same parameters as $G_{\perp}(k)$ in Fig.~\ref{corrtra_fit}.
The simulated data points are shown by circles, the error bars being indicated,
where these exceed the symbol size.
The fits to~(\ref{eq:appar}) within $k < 0.28$ and $k < 0.55$ are shown by
solid and dashed curves, respectively. The upper values of the fit intervals
are indicated by vertical dashed lines.}
\label{corrlong_fit}
\end{figure}

In Figs.~\ref{corrtra_fit} and~\ref{corrlong_fit}, some of our fit curves at $h=0.0004375$ are shown,
which are relatively good, especially for $n=4$ and $n=10$.

\section{Test of the Patashinski--Pokrovski relation}
\label{sec:PP}

Since the fit curves for the $O(4)$ and $O(10)$ models in Figs.~\ref{corrtra_fit} 
and~\ref{corrlong_fit} provide good approximations for the correlation functions in the
thermodynamic limit at the given parameters and small $k$ values, we have applied these
analytic approximations to test the PP relation~(\ref{eq:PP}) for large distances $x$.
We have used Eqs.~(\ref{eq:Gx}) and~(\ref{eq:cutoff}) for this purpose.
In the case of a finite lattice, the wave vectors belong to a cube with 
$-\pi \le k_x \le \pi$, $-\pi \le k_y \le \pi$, $-\pi \le k_z \le \pi$. Therefore
a reasonable choice of the cut-off parameter is $\Lambda = \pi$. The precise value of
$\Lambda$, however, is not important, since the result for $\tilde{G}_i(x)$ is 
insensitive to the variation of $\Lambda$ at large enough $x$. 
We calculate functions $f_{\perp}(x)$ and $f_{\parallel}(x)$ given by
\begin{eqnarray}
f_{\perp}(x) &=& 2 \ln \left(x \, \tilde{G}_{\perp}(x) \right) + \ln \left( \frac{n-1}{2M^2} \right) \;, \label{fper} \\
f_{\parallel}(x) &=& \ln \left( x^2 \, \tilde{G}_{\parallel}(x) \right) \label{fpar} \;,
\end{eqnarray}
which have to be equal if the PP relation holds. In the Gaussian 
approximation~(\ref{eq:GGauss}) we have $\tilde{G}_{\perp}(x) \propto e^{-mx}/x$ at $x \to \infty$,
implying the linearity of these functions at large $x$. 

\begin{figure}
\begin{center}
\includegraphics[width=0.5\textwidth]{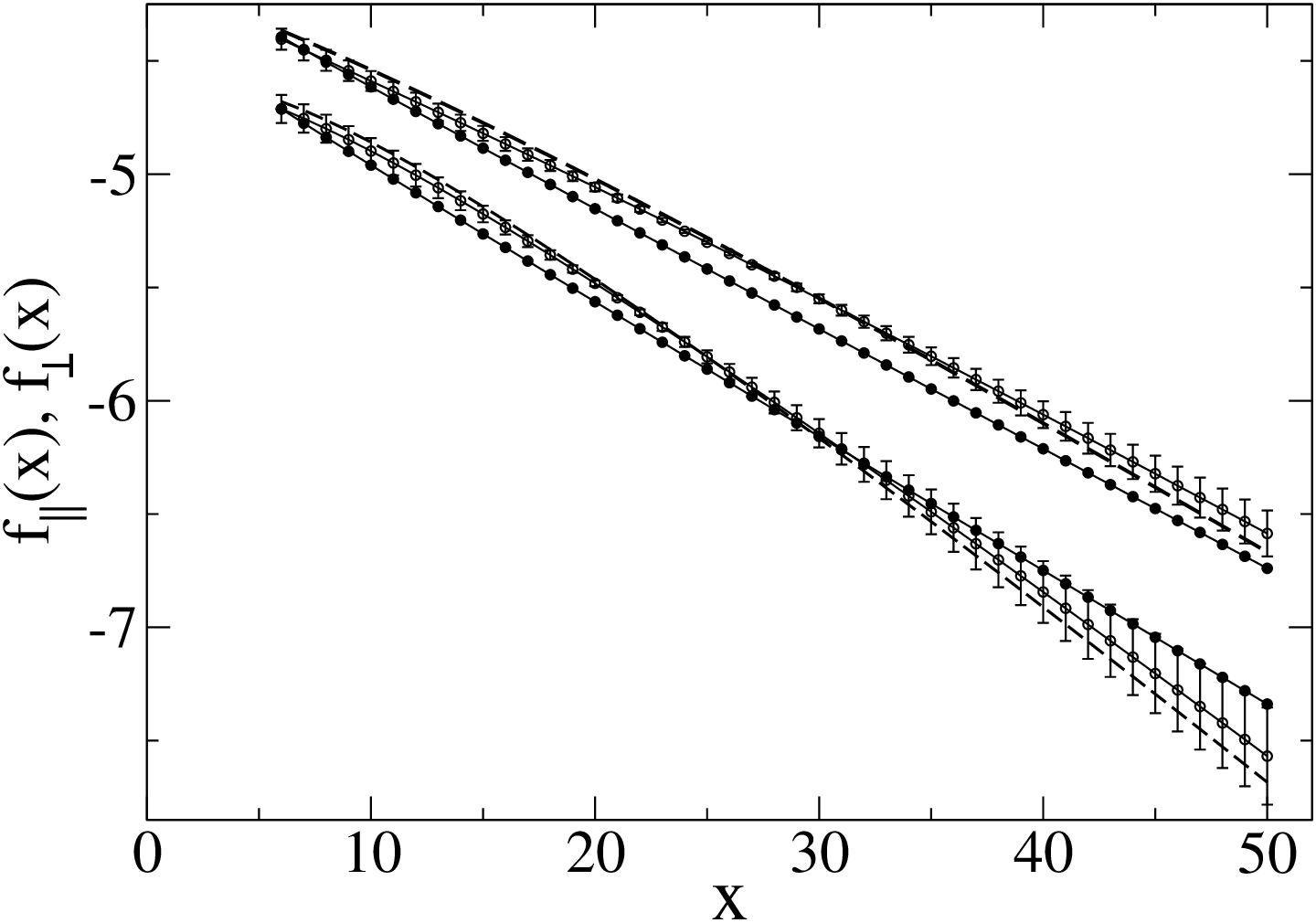}
\end{center}
\caption{The functions 
$f_{\perp}(x)$~(\ref{fper}) and $f_{\parallel}(x)$~(\ref{fpar})
for the $O(4)$ (upper curves) and $O(10)$ (lower curves) models, evaluated from~(\ref{eq:Gx})
(replacing $i$ with $\perp$ or $\parallel$) and~(\ref{eq:cutoff}) with $\Lambda= \pi$, 
using the fit functions $G_{\perp}(k)$ and $G_{\parallel}(k)$ in Figs.~\ref{corrtra_fit} and~\ref{corrlong_fit}. 
The solid circles correspond to $f_{\perp}(x)$, the error bars being smaller than the symbol size. The empty circles 
with error bars represent $f_{\parallel}(x)$, evaluated using the fit interval $k<0.28$ in Fig.~\ref{corrlong_fit},
the results for $k<0.55$ being indicated by dashed curves. 
The PP relation~(\ref{eq:PP}) implies $f_{\perp}(x)=f_{\parallel}(x)$.}
\label{pptest}
\end{figure}

The magnetization $M$ for the actual parameters are taken from~\cite{KMR10,KMR12}. We have considered
the distances $x \ge 6$, as in this case the $f_{\parallel}(x)$ curves at $\Lambda = \pi$
and $\Lambda = \pi/2$ lie practically on top of each other. The function $f_{\perp}(x)$ 
is even much less influenced by the change of $\Lambda$. 
The used here fits in Fig.~\ref{corrtra_fit} are perfect, whereas those in Fig~\ref{corrlong_fit}
show some systematic variations depending on the fit interval. The fits over $k<0.28$
are better for small $k$, therefore they could provide a better approximation of
$f_{\parallel}(x)$ for large $x$, although the fits over
a wider interval $k<0.55$ look better on average. We have compared the results in both
cases to judge about the magnitude of systematic errors. The resulting curves
of $f_{\perp}(x)$ and $f_{\parallel}(x)$ within $6 \le x \le 50$ are shown in Fig.~\ref{pptest}.
The errors due to statistical and systematic uncertainties in the fit parameters increase significantly
for $x > 50$, therefore no larger distances are considered here.  
As we can judge from Fig.~\ref{pptest}, the PP relation holds approximately (within 10\% or 15\% accuracy)
in these examples at a finite external field $h=0.0004375$. 

Another case, where the PP relation can be tested, is the large--$x$ behavior in the thermodynamic
limit at $h=+0$. It is closely related to the universal ratio test in Sec.~\ref{sec:ratios}.
The PP relation states that~(\ref{eq:st1}) must hold for the ratio $B M^2/A^2$. As it is shown in Sec.~\ref{sec:ratios},
this requirement is equivalent to~(\ref{eq:st2}) for the ratio $b M^2/a^2$, if the transverse exponent
is $\lambda_{\perp}=2$, as predicted by the standard theory. Tests
in Sec.~\ref{sec:ratios} show certain inconsistencies with~(\ref{eq:st2}) (see Fig.~\ref{R0}) 
and, consequently, with the PP relation if $\lambda_{\perp}=2$. Assuming that $\lambda_{\perp}=2$
holds at $n=10$, the ratio $B M^2/A^2=8 bM^2/a^2=3.984 \pm 0.080$ (see Eq.~(\ref{eq:ratiosst})) appears to be somewhat 
smaller than the value $4.5$ expected
from the PP relation. One can use~(\ref{eq:ratios}) to calculate $B M^2/A^2$ from $b M^2/a^2$ at our
numerically estimated values of the exponent 
$\lambda_{\perp}<2$. It leads to slightly (by $\sim 1 \%$)
smaller values of $B M^2/A^2$. Thus, we find that the PP relation holds approximately
(within about $12 \%$ accuracy in our examples) in the thermodynamic limit for large $x \to \infty$ at $h=+0$.

\section{Conclusions}

In the current paper, we have considered the behavior of the longitudinal and transverse correlation
functions and Goldstone mode singularities in $O(n)$ models from different aspects compared to our earlier
Monte Carlo studies~\cite{KMR08,KMR10,K2012,KMR12}. Apart from the two--point correlation functions, here
we have calculated the two--plane correlation functions, which are very important for the provided here
discussions related to the recent work by Engels and Vogt~\cite{EV}. 
We confirm the stated in~\cite{EV} fact that the transverse two--plane correlation function
of the $O(4)$ model for lattice sizes about $L=120$ and small 
external fields $h$ is very well described by a Gaussian approximation with $\lambda_{\perp}=2$ in~(\ref{a}). 
However, we have shown in Sec.~\ref{sec:fitstwo} that fits of not lower quality are provided by certain non--Gaussian
approximation, where $\lambda_{\perp}<2$. Thus, the behavior of the two--plane correlation functions does not imply 
that the $O(4)$ model is essentially Gaussian with $\lambda_{\perp}=2$.
We have also tested the cases $n=2, 4, 10$ for larger lattice
sizes (e.~g., $L=350$ and $L=512$), where not as good agreement with the
Gaussian model has been observed.

The ratio $b M^2/a^2$ has been considered in Sec.~\ref{sec:ratios}, showing that its universality
follows not only from the GFD theory~\cite{K2010}, but also from the standard theory, yielding
$b M^2/a^2 = (n-1)/16$. Our MC estimates of this ratio are $0.06 \pm 0.01$ for $n=2$, $0.17 \pm 0.01$ for $n=4$ 
and $0.498 \pm 0.010$ for $n=10$. The latter estimate shows a very remarkable, as compared
to the error bars, deviation from the standard--theoretical value $9/16=0.5625$.
Our MC estimation in~\cite{KMR08,KMR10,KMR12} points to small deviations from the standard--theoretical
predictions in favor of the GFD theory. A clear evidence that the standard theory is not asymptotically 
exact (as one often claims) 
at large length scales has been provided in~\cite{K2012}, showing that a self consistent 
(within the standard theory) estimation of the longitudinal exponent 
$\lambda_{\parallel}$ from MC data of the three--dimensional $O(2)$ model at $\beta = 0.55 > \beta_c$ yields
 $\lambda_{\parallel}= 0.69 \pm 0.10$ in disagreement with the expected value $\lambda_{\parallel}= 1$.
The current MC estimation of the ratio $b M^2/a^2$ provides one more such evidence. 

In Sec.~\ref{sec:analap}, we have proposed and tested certain analytic approximations for 
the two--point correlation functions $G_{\perp}(k)$ and $G_{\parallel}(k)$ in $\langle 100 \rangle$ direction
and also for $G_{\perp}({\bf k})$ and $G_{\parallel}({\bf k})$ at small $k = \mid {\bf k} \mid$, 
which are consistent with the expected behavior at $h=+0$ and are valid also at a finite external field $h$. 
We have found that these approximations (Eqs.~(\ref{eq:apper}) and~(\ref{eq:appar})) fit reasonably well 
the simulation data for small $k$. The exponents $\lambda_{\perp}$ and $\lambda_{\parallel}$
in~(\ref{eq:apper})--(\ref{eq:appar}) have been discussed as fit parameters, showing that these
are comparable with our earlier estimates. In Sec.~\ref{sec:PP}, we have used our analytic approximations
to test the Patashinski--Pokrovski relation~(\ref{eq:PP}), and have found that it holds approximately
within the accuracy of about $10\%$ or $15\%$ in the examples considered.

\section*{Acknowledgments}

This work was made possible by the facilities of the
Shared Hierarchical Academic Research Computing Network
(SHARCNET:www.sharcnet.ca). 
R. M. acknowledges the support from the
NSERC and CRC program.

\end{document}